%% file: main.tex
\documentclass[twocolumn,twocolappendix]{aastex63}

\graphicspath{{./}{figures/}}
\usepackage[nolist]{acronym}
\input{acronyms.tex}
\usepackage{breqn}
\usepackage{hyperref}

\newcommand{\citeac}[2]{\aclu{#1} \citep[\ac{#1};][]{#2}}

\newcommand{\mujysb}{$\mu$$\rm Jy$$\,/\, arcsec^2$}

\begin{document}
\title{JWST/NIRCam Detection of the Fomalhaut C Debris Disk in Scattered Light}

\correspondingauthor{Kellen Lawson}
\email{kellenlawson@gmail.com}
\author[0000-0002-6964-8732]{Kellen Lawson}
\affiliation{NASA-Goddard Space Flight Center, Greenbelt, MD, USA}
\affiliation{NASA Postdoctoral Program Fellow}

\author[0000-0001-5347-7062]{Joshua E. Schlieder}
\affiliation{NASA-Goddard Space Flight Center, Greenbelt, MD, USA}

\author[0000-0002-0834-6140]{Jarron M. Leisenring}
\affiliation{Steward Observatory, University of Arizona, 933 N. Cherry Avenue, Tucson, AZ 85721, USA}

\author[0000-0002-7325-5990]{Ell Bogat}
\affiliation{NASA-Goddard Space Flight Center, Greenbelt, MD, USA}
\affiliation{Department of Astronomy, University of Maryland, College Park, MD 20782, USA}

\author[0000-0002-5627-5471]{Charles A. Beichman}
\affiliation{NASA Exoplanet Science Institute/IPAC, Jet Propulsion Laboratory, California Institute of Technology, 1200 E California Blvd, Pasadena, CA 91125, USA}

\author[0000-0001-5966-837X]{Geoffrey Bryden}
\affiliation{Jet Propulsion Laboratory, California Institute of Technology, Pasadena, CA, USA}

\author[0000-0001-8612-3236]{Andr\'as G\'asp\'ar}
\affiliation{Steward Observatory, University of Arizona, 933 N. Cherry Avenue, Tucson, AZ 85721, USA}

\author[0000-0001-5978-3247]{Tyler D. Groff}
\affiliation{NASA-Goddard Space Flight Center, Greenbelt, MD, USA}

\author[0000-0003-0241-8956]{Michael W. McElwain}
\affiliation{NASA-Goddard Space Flight Center, Greenbelt, MD, USA}

\author[0000-0003-1227-3084]{Michael R. Meyer}
\affiliation{Department of Astronomy, University of Michigan, 1085 S. University, Ann Arbor, MI 48109, USA}

\author[0000-0001-7139-2724]{Thomas Barclay}
\affiliation{NASA-Goddard Space Flight Center, Greenbelt, MD, USA}

\author[0000-0002-5335-0616]{Per Calissendorff}
\affiliation{Department of Astronomy, University of Michigan, 1085 S. University, Ann Arbor, MI 48109, USA}

\author[0000-0003-1863-4960]{Matthew De Furio}
\affiliation{Department of Astronomy, University of Michigan, 1085 S. University, Ann Arbor, MI 48109, USA}

\author[0000-0002-6845-9702]{Yiting Li}
\affiliation{Department of Astronomy, University of Michigan, 1085 S. University, Ann Arbor, MI 48109, USA}

\author[0000-0002-7893-6170]{Marcia J. Rieke}
\affiliation{Steward Observatory, University of Arizona, 933 N. Cherry Avenue, Tucson, AZ 85721, USA}

\author[0000-0001-7591-2731]{Marie Ygouf}
\affiliation{Jet Propulsion Laboratory, California Institute of Technology, Pasadena, CA, USA}

\author[0000-0002-8963-8056]{Thomas P. Greene} \affiliation{NASA Ames Research Center, MS 245-6, Moffett Field, CA 94035, USA}

\author[0000-0001-8627-0404]{Julien H. Girard}
\affiliation{Space Telescope Science Institute, 3700 San Martin Drive, Baltimore, MD 21218, USA}

\author[0000-0002-5581-2896]{Mario Gennaro}
\affiliation{Space Telescope Science Institute, 3700 San Martin Drive, Baltimore, MD 21218, USA}
\affiliation{The William H. Miller {\sc III} Department of Physics \& Astronomy, Bloomberg Center for Physics and Astronomy, Johns Hopkins University, 3400 N. Charles Street, Baltimore, MD 21218, USA}

\author[0000-0003-2769-0438]{Jens Kammerer}
\affiliation{European Southern Observatory, Karl-Schwarzschild-Straße 2, 85748 Garching, Germany}

\author[0000-0002-4410-5387]{Armin Rest}
\affiliation{Space Telescope Science Institute, 3700 San Martin Drive, Baltimore, MD 21218, USA}
\affiliation{The William H. Miller {\sc III} Department of Physics \& Astronomy, Bloomberg Center for Physics and Astronomy, Johns Hopkins University, 3400 N. Charles Street, Baltimore, MD 21218, USA}

\author[0000-0002-6730-5410]{Thomas L. Roellig}
\affiliation{NASA Ames Research Center, MS 245-6, Moffett Field, CA 94035, USA}

\author[0000-0003-3759-8707]{Ben Sunnquist}
\affiliation{Space Telescope Science Institute, 3700 San Martin Drive, Baltimore, MD 21218, USA}

\submitjournal{\apjl}
\accepted{2024 April 28}

\begin{abstract}
Observations of debris disks offer important insights into the formation and evolution of planetary systems. Though M dwarfs make up {approximately $80 \%$ of nearby stars}, very few M-dwarf debris disks have been studied in detail — making it unclear how or if the information gleaned from studying debris disks around more massive stars extends to the more abundant M dwarf systems. We report the first scattered-light detection of the debris disk around the M4 star Fomalhaut C using JWST’s Near Infrared Camera (NIRCam; 3.6$~\micron{}$ and 4.4$~\micron{}$). 
This result adds to the prior sample of only four M-dwarf debris disks with detections in scattered light, and marks the latest spectral type {and oldest } star among them. The size and orientation of the disk in these data are generally consistent with the prior ALMA sub-mm detection. Though no companions are identified, these data provide strong constraints on their presence — with sensitivity sufficient to recover sub-Saturn mass objects in the vicinity of the disk. This result illustrates the unique capability of JWST for uncovering elusive M-dwarf debris disks in scattered light, and lays the groundwork for deeper studies of such objects in the 2–5$~\micron$ regime.
\end{abstract}

\keywords{}

\section{Introduction}\acresetall

Studies of debris disks, the dust-dominated remnants of the planet formation process, offer a number of compelling insights into how planetary systems form and evolve. Analysis of debris disk morphology provides a window into the dynamical history of these systems — revealing past stellar interactions \citep[e.g.,][]{Ardila2005} or uncovering the ``sign-posts" of yet-unseen companions \citep[e.g.,][]{Kalas2005,Kenyon2008}. Observational signatures in disks — such as sharp edges and gaps, warps, kinks, and filaments, brightness asymmetries, dust clumps, and other deviations from a smooth radial profile — have all been used to suggest the existence and properties of potential planets \citep[e.g.,][]{Hashimoto2012, Gaspar2023}{. Simulations of these disk signatures have shown that they might be explained by direct planetary interactions \citep[e.g.,][]{Dong2012, Lee2016}, indirect planetary interactions \citep[e.g.,][]{Yelverton2018}, or} other phenomena relevant to understanding planet formation processes \citep[e.g.,][]{Dong2018}. Indeed, such morphological signatures have successfully predicted the presence of previously-unseen planets \citep[e.g., $\beta$ Pictoris b;][]{Lagrange2009}. Analysis of debris disks' brightness as a function of wavelength via multi-wavelength studies enables assessment of the composition and size distribution of the constituent dust \citep{Rodigas2015, Ballering2016, Arnold2019}. This provides the opportunity to study the composition of planetesimals left over from the formation of planets, while also helping us to assess the mechanisms driving the evolution of post-accretion circumstellar material \citep[e.g.,][]{Meyer2007}.

Though M dwarfs make up {approximately 80\% of nearby stars \citep[][]{Reyle2021}}, few M-dwarf debris disks have been identified, and fewer studied in detail \citep[e.g.,][]{Avenhaus2012, Luppe2020, Ren2021, Cronin-Coltsmann2023}. Of the 10 {known M-dwarf debris disk systems within 100 pc}, only five have had their disks definitively spatially resolved \citep{Cronin-Coltsmann2023}{, and only four have been detected in scattered light. This is in contrast} to the $\sim 46$ resolved debris disks around A-type stars\footnote{Based on the catalog of resolved disks at \href{https://www.circumstellardisks.org/index.php?catalog=resolved}{circumstellardisks.org}}. The small sample of M-dwarf debris disks available for study {precludes assessment of how the debris disks of the low-mass majority of stars, and thus their planetary systems, differ from those of more massive stars.} While numerous studies have proposed mechanisms that might lead to M-dwarf debris disks being intrinsically less common than those of higher-mass stars \citep[e.g.,][]{Lestrade2011}, recent results suggest that they are equally abundant, but are often simply too faint to detect \citep{Luppe2020, Cronin-Coltsmann2023}. Moreover, simply detecting an M-dwarf debris disk is not sufficient for accessing the wealth of information that debris disks can provide. Deep morphological studies require strong, well-resolved detections to meaningfully disentangle dynamical scenarios{. For example, precise measurements of disk eccentricity can test the possibility of past stellar interactions \citep[e.g.,][]{Cronin-Coltsmann2021}, while precise measurement of spiral arm motion can distinguish between their likely drivers — either gravitational instability or a planetary perturber \citep[e.g.,][]{Dong2018}}. Meanwhile, studies of debris disk composition and dust properties require detections at multiple wavelengths — spanning both scattered light and thermal emission — to break degeneracies \citep[e.g.,][]{Rodigas2015, Ballering2016}. In both cases, the difficulty of resolving M-dwarf debris disks introduces significant barriers for understanding their planetary systems.

Of the five well-resolved M-dwarf debris disks, four have been detected in scattered-light: AU Mic \citep{Kalas2004}, TWA 7 \citep{Choquet2016}, TWA 25 \citep{Choquet2016}, and GSC 07396-00759 \citep{Sissa2018}. This leaves one system which has been previously resolved only in thermal emission: Fomalhaut C \citep[M4, age $\sim440$ Myr, $d=7.7$ pc;][]{Mamajek2013, GaiaDR32022, Cronin-Coltsmann2021}. Following the Fomalhaut C disk's initial discovery based on Herschel photometry \citep{Kennedy2014}, 870$~\micron$ observations from ALMA successful resolved the moderately inclined ($\sim 44 \degr$) debris disk for the first time \citep[][]{Cronin-Coltsmann2021}. However, efforts to detect the disk in visible and near-IR scattered light with HST/STIS and VLT/SPHERE resulted in non-detections \citep{Cronin-Coltsmann2021}. 

JWST's \ac{nircam} is a stable space-based high-contrast imager operating at \ac{nir} wavelengths \citep[0.6–5~\micron,][]{Rieke2005}. Since light scattered by dust around intrinsically redder M dwarfs can result in higher surface brightness in the near-IR than at optical wavelengths \citep[even for disks with intrinsically blue scattering; e.g.,][]{Ren2021}, observations with NIRCam can ease the requisite sensitivity for detection of M-dwarf debris disks in scattered light. Additionally, \ac{nircam} provides the capability to characterize debris disks in the $2-5~\micron$ range \citep[e.g.,][]{Lawson2023}, where ground-based observations are limited by telluric absorption and {background levels}. {NIRCam's ability to use broader filters than the common ground-based filters at these wavelengths may likewise provide an advantage for efforts to study these disks.} With many important scattered-light spectral features occurring at these wavelengths \citep[e.g., the 3~$\micron$ water-ice feature;][]{Kim2019}, detections here are particularly powerful for diagnosing debris disk composition. Combined with the utility of ALMA for providing thermal detections of these objects \citep{Luppe2020, Cronin-Coltsmann2023}, JWST detections in scattered-light may provide a compelling new avenue for understanding the planetary systems of M dwarfs.

\citet{Lawson2023} demonstrated the efficacy of \ac{nircam} coronagraphy for studying M dwarf debris disks at 3–5$~\micron$ in application to the debris disk of AU Mic. Herein, we build upon these results to present the first scattered-light detections of the Fomalhaut C debris disk using coronagraphic observations from JWST/NIRCam in F356W (3.6~\micron{}) and F444W (4.4~\micron{}). Based on these data, we provide analysis of the disk's scattered-light morphology, brightness, and color, and also conduct a deep search for companions in the data — reaching sensitivities sufficient to detect masses below that of Saturn.

\section{Observations}\label{sec:observations}
Fomalhaut C was observed as part of a JWST NIRCam Guaranteed Time Observing (GTO) program. This program, GTO 1184 (PI J. Schlieder)\footnote{\href{https://www.stsci.edu/jwst/science-execution/program-information.html?id=1184}{GTO 1184 - A NIRCam Coronagraphic Imaging Survey of Nearby Young M Dwarfs}}, targeted 9 nearby, young M dwarfs with NIRCam coronagraphy to search for low-mass companions. A full description of the GTO 1184 survey and its results will be presented in a separate paper (Bogat et al.~in prep).

All targets were observed at two roll angles separated by $\sim$$10\degr$ using two filters from the NIRCam long wavelength (LW) channel (average pixel scale of 63 mas$/$pixel): F356W ($\lambda_{pivot} = 3.57 \, \micron$, $\Delta \lambda = 0.77 \, \micron$) and F444W ($\lambda_{pivot} = 4.36 \, \micron$, $\Delta \lambda = 0.94 \, \micron$)\citep{Rieke2005, Leisenring2021}. Observing at two roll angles not only permits the use of \citeac{adi}{Marois2006} for companion searches, but also enables any identified companions or small-scale disk features ({which are fixed to the sky frame}) to be distinguished from residual speckle noise or diffraction artifacts ({which are fixed to the detector frame}).
The \texttt{MASK335R} coronagraph mask, which has an inner working angle (IWA)\footnote{Where the IWA is defined as the angular separation at which
coronagraph transmission reaches 50\%} of $0\farcs63$, was used for observations in both filters \citep{Krist2009}. Using the \texttt{SUB320} subarray mode, each integration results in an image of 320$\times$320 pixels (20$\times$20$\arcsec$).

Instead of observing dedicated reference stars for each target, the survey permits \ac{rdi} using a self-referencing strategy — wherein \ac{rdi} is performed for each target using the other survey targets as a reference library. The reference library that we utilize for Fomalhaut C includes four GTO 1184 survey targets found to be free of (detectable) circumstellar flux: AP Col, 2MJ0944, G-7-34, and HIP 17695. These observations are summarized in Table \ref{tab:observations}. 

Of the reference targets, AP Col and 2MJ0944 were observed very close in time to the observations of Fomalhaut C, with no intervening tilt events or wavefront corrections. The remaining reference targets, G-7-34 and HIP 17695, were observed approximately two months prior to Fomalhaut C. During this interval, one large and one moderate tilt event occurred \citep{Lajoie2023}. As such, these targets' images are unlikely to improve \ac{psf} subtraction at separations where speckle noise dominates, but may nevertheless help to suppress noise in the background-limited regime.

\begin{deluxetable*}{@{\extracolsep{2pt}}ccccccccc}
    \tablewidth{0pt}
    \tablecaption{JWST/NIRCam Observations}
    \tablehead{
    \colhead{} & \colhead{} & \colhead{} & \colhead{} & \colhead{} & \multicolumn{2}{c}{Roll 1 Offset\tablenotemark{c}} & \multicolumn{2}{c}{Roll 2 Offset\tablenotemark{c}} \\
    \cline{6-7}
    \cline{8-9}
    \colhead{Target ID\tablenotemark{a}} & \colhead{Spec. Type} & \colhead{W1$\,$(mag)\tablenotemark{b}} & \colhead{W2$\,$(mag)\tablenotemark{b}} & \colhead{Obs. Date} & \colhead{$\rm \Delta X$ (mas)} & \colhead{$\rm \Delta Y$ (mas)} & \colhead{$\rm \Delta X$ (mas)} & \colhead{$\rm \Delta Y$ (mas)}
    }
    \startdata
    HIP 17695 & M3 & 6.81 & 6.66 & 2022 Oct 3 & 8.5 & -5.2 & 10.5 & 5.6 \\
    G-7-34 & M4 & 8.01 & 7.81 & 2022 Oct 3 & 8.8 & 4.9 & 11.0 & 0.6 \\
    FOMALHAUT-C & M4 & 6.92 & 6.79 & 2022 Nov 26 & -13.5 & 13.3 & 21.4 & -5.0 \\
    V-AP-COL & M4.5 & 6.67 & 6.39 & 2022 Nov 27 & 3.2 & -8.3 & 14.2 & -7.9 \\
    2MJ0944\tablenotemark{d} & M5 & 7.36 & 7.19 & 2022 Nov 27 & 3.9 & 0.6 & -1.2 & 5.9 \\
    \enddata
    \tablecomments{A summary of the science and reference targets and their observations used in this study. All targets except Fomalhaut C (``FOMALHAUT-C") were used as reference targets. All exposures used the \texttt{MEDIUM8} readout pattern with 10 groups per integration and with 8 integrations per roll for F356W and 16 integrations per roll for F444W — for total exposure times of 1676 seconds and 3562 seconds respectively. \tablenotetext{a}{The identifier used for each target in the GTO 1184 program.} \tablenotetext{b}{ALLWISE photometry; \citet{Wright2010}} \tablenotetext{c}{The measured offsets between the target star and coronagraph center in detector coordinates (see Section \ref{sec:reduction}).} \tablenotetext{d}{2MASS J09445422-1220544}}
    \label{tab:observations}
\end{deluxetable*}

\section{Data Reduction}\label{sec:reduction}
We reduce the data predominantly following the procedure of \citet{Lawson2023}, but also adopting a number of more recent additions to the \texttt{spaceKLIP} package \citep{Kammerer2022}\footnote{\texttt{spaceKLIP} 
 version \href{https://github.com/kammerje/spaceKLIP/tree/f64258dbb48d908a2f668eb2e073f9351bf646fe}{\texttt{1.0.1.dev185+gf64258d}}.} and other changes, as summarized hereafter. 

Using the \texttt{rampfit} step from \texttt{spaceKLIP}, we process Stage 0 products (\textit{*uncal.fits}) to Stage 1 (\textit{*rateints.fits}). For this purpose, we skip dark current subtraction\footnote{As described in \citet{Carter2022}, the dark reference files specific to the coronagraphic subarrays in NIRCam have insufficient signal-to-noise ratio to be useful in this context.}, adopt a border of 4 pixels as pseudo reference pixels, and use a jump detection threshold of 4. This step now also includes a 1$\,$/$\,$f {(frequency)} noise correction procedure to eliminate instrumental striping that can otherwise manifest in the data {\citep[see description in][]{Rebollido2024}}.

Using the \texttt{imgprocess} step from \texttt{spaceKLIP}, we then process the Stage 1 products to Stage 2 (\textit{*calints.fits}), including identification of outlier pixels with the outlier detection step. Next, we identify additional bad pixels and correct them (using \texttt{fix\_bad\_pixels} and \texttt{replace\_nans} from \texttt{spaceKLIP.imagetools}). Following this, manual inspection revealed six additional pixels with seemingly spurious values throughout the data {(likely unflagged static hot pixels)}. Each of these pixels was replaced with the median value within a 5$\times$5 pixel box.

After correction of outlier pixels, we fit and subtract a background model from each science and reference exposure. In this model, the distribution of background flux in the coronagraphic data is simulated by multiplying a uniform image by the coronagraph transmission map (including the neutral density squares; generated using \texttt{WebbPSF\_ext}, \citealt{Leisenring2021}) and then convolving the result with a grid of spatially varying PSFs from \texttt{WebbPSF\_ext} — following the general convolution procedure described in \citet[][Appendix A]{Lawson2023}. Justification of the need for this step and the details of our implementation are provided in Appendix \ref{app:bg_models}.

Following background subtraction, image alignment is carried out as described in \citet{Lawson2023} — using a synthetic stellar PSF to determine the positions of the occulted star in each exposure. The results of this process (see Table \ref{tab:observations}) show that the Fomalhaut C target acquisition (TA) procedure incurred particularly large errors, resulting in significant misalignment between the star and the coronagraph. As none of the available reference targets had similar misalignment, this should be expected to result in more significant speckle residuals at small separations following PSF-subtraction — inhibiting the sensitivity of the data in the speckle-dominated regime.

\section{Reference Star Differential Imaging}\label{sec:psfsub}
To remove the pattern of diffracted starlight in each image, we adopt three distinct strategies: classical \ac{rdi}, model constrained RDI \citep[MCRDI,][]{Lawson2022,Lawson2023}, and a variation of synthetic RDI \citep[SRDI,][]{Greenbaum2023}. In each case, a final image is created by derotating and median combining the starlight-subtracted integrations (\textit{*calints.fits} files) across both rolls.

\subsection{Classical RDI}
First, the classical RDI procedure is carried out for each filter by median combining the integrations of AP Col (``V-AP-COL''), which is the better spectral match of the two GTO 1184 targets observed near-contemporaneously with Fomalhaut C (see Table \ref{tab:observations}). We then scale the brightness of the PSF model to minimize squared residuals with the data within a 20 pixel radius aperture centered on the star (chosen to exclude the disk signal as identified in \citealt{Cronin-Coltsmann2021}). We find no visual improvement in the subtraction by manually tuning the determined scaling factor by small amounts in either direction. 

\subsection{Model Constrained RDI}
In Model Constrained RDI (MCRDI), a synthetic image of the circumstellar scene is used to prevent systematic overestimation of the stellar PSF's brightness during construction of the stellar PSF model \citep{Lawson2022, Lawson2023, Rebollido2024}, which otherwise results in so-called ``oversubtraction" \citep[e.g.,][]{Pueyo2016}. To apply MCRDI, an initial unconstrained RDI reduction is carried out. The RDI residuals are then fit with disk models using standard forward modeling techniques to identify the best-fitting disk model image. The resulting disk model is then used to carry out an MCRDI reduction of the data {\citep{Lawson2022}. In the MCRDI procedure, the optimal stellar PSF model is constructed by comparing the reference images to science images from which the best-fitting disk model has been subtracted (where, in our implementation, the PSF model is formed from a linear combination of reference images). The resulting stellar PSF model is then subtracted from the original science images. This technique provides final images which are effectively free of oversubtraction and enables more accurate disk photometry than application of model-based photometric throughput corrections \citep{Lawson2022}.} Since the F356W data have better spatial resolution than the F444W data, we first model the disk at F356W and then restrict the F444W model geometry to match the geometry of the F356W result, while still varying the scattering phase function and overall brightness of the disk. Descriptions of the disk models, model convolution, the PSF subtraction algorithm settings, and handling of background sources are provided in Appendix \ref{app:mcrdi_details}.

\subsection{Synthetic RDI}
Variation in the on-sky NIRCam PSF is predominantly a function of a) the alignment between the target star and the coronagraph mask, b) the evolving wavefront error, and c) the spectral type of the star \citep[e.g.,][]{Girard2022}. The \texttt{WebbPSF} tool \citep{Perrin2014} provides the ability to generate synthetic NIRCam PSFs where these factors are accounted for by (respectively) tuning mask alignment, using empirical optical path difference (OPD) maps, and providing a source spectrum. In some scenarios, these synthetic PSFs could be used in place of or to supplement on-sky reference PSFs for carrying out \ac{rdi} \citep[see, e.g.,][]{Greenbaum2023}. This is a particularly appealing prospect in this application, given the previously noted differences in the coronagraph alignment between the science and reference data. 

Though providing the aforementioned options for tuning of synthetic PSF models, the resulting models from \texttt{WebbPSF} are nevertheless imperfect. This is both because of imperfect knowledge of the true state of these parameters (mask alignment, wavefront error, and target spectrum), and because of inaccuracies or simplifications in the underlying \texttt{WebbPSF} optical model. In the case of the Fomalhaut C data, performing \ac{rdi} with a nominal \texttt{WebbPSF} model (using the nearest available OPD file to the observations, the measured mask offsets, and an approximate AMES-Cond stellar spectrum, \citealt{Allard2001}) leaves significant, bright PSF residuals well in excess of the residuals from classical RDI or MCRDI. 

To improve the nominal synthetic PSF models, we proceed with a strategy of empirical model corrections — leveraging the available reference images to mitigate the recurring inaccuracies in the synthetic PSFs. This procedure is detailed in Appendix \ref{app:synth_rdi}. By introducing these corrections, some noise is added to the otherwise noiseless synthetic PSF models. However, contrasts show a net improvement of up to a factor of six — reaching sensitivities comparable to those of MCRDI.

\section{Results}\label{sec:results}

\subsection{Detection of the Fomalhaut C Debris Disk in Scattered Light}\label{sec:detection}
\begin{figure*}
\centering
\includegraphics[width=0.99\textwidth]{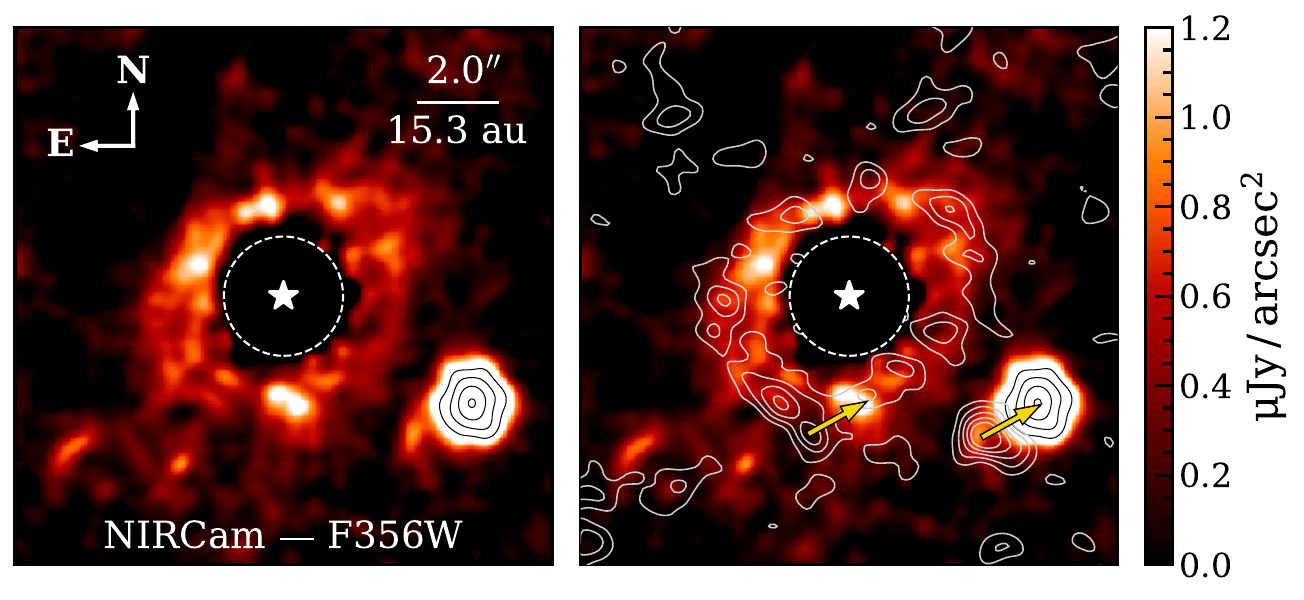}
\caption{Left: NIRCam F356W detection of the Fomalhaut C debris disk in scattered light resulting from the MCRDI procedure (Section \ref{sec:psfsub}). The image has been smoothed by a gaussian kernel with FWHM of 6.7 pixels ($3\times$ the PSF FWHM). The inner software mask has a radius of 1$\farcs$5 and covers the region of significant residual speckle noise. Logarithmically-spaced black contours are drawn above the maximum for the color stretch to show the peak location of the bright south-western background source. Right: As the image on the left, but with silver contours showing the 1–5$\,\sigma$ levels for the ALMA detection of the disk reported in \citet{Cronin-Coltsmann2021}. The gold arrows extend from the locations of the two assumed background sources modeled in \citet{Cronin-Coltsmann2021}, showing their expected displacement due to the proper motion of Fomalhaut C \citep{GaiaDR32022}. Both predicted positions are consistent with sources detected in the 2022 NIRCam data, supporting their identity as unassociated background objects.\label{fig:fomc_with_pm}}
\end{figure*}

\begin{figure*}[h]
\centering
\includegraphics[width=0.98\textwidth]{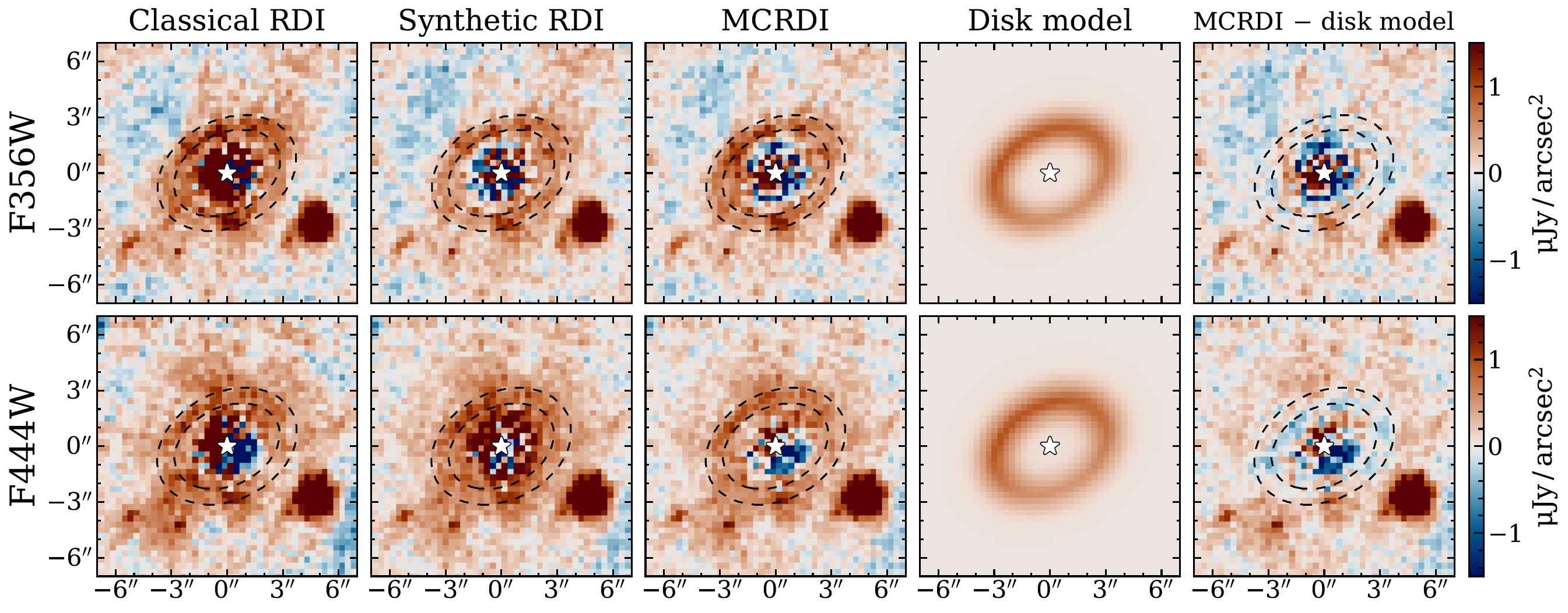}
\caption{PSF-subtracted final images of Fomalhaut C in F356W (top row) and F444W (bottom row), as well as the MCRDI disk model and disk model residuals (two rightmost columns). To reduce noise, each image has been median binned by a factor of 5 (such that each pixel in the binned image is the median of a $5\times5$ pixel square in the original image). The first three columns correspond to the three PSF subtraction strategies (Section \ref{sec:psfsub}): classical RDI, synthetic RDI, and model constrained RDI (MCRDI), respectively. The dashed ellipses mark the approximate shape and size of the disk identified by \citet{Cronin-Coltsmann2021}.\label{fig:fomc_detections}}
\end{figure*}

In both F356W and F444W, and using all three starlight-subtraction techniques, faint extended signal is detected in the vicinity of Fomalhaut C (Figures \ref{fig:fomc_with_pm}, \ref{fig:fomc_detections}). Though manifesting at marginal signal-to-noise per resolution element (SNRE$\;\sim 3$ in F356W), the distribution of the flux is generally consistent with the orientation and size of the Fomalhaut C disk as seen in the ALMA detection reported in \citet{Cronin-Coltsmann2021} (indicated by the silver contours in Figure \ref{fig:fomc_with_pm} and by the dashed ellipses in Figure \ref{fig:fomc_detections}). For F444W, additional flux is seen extending to wider separations; this is discussed further in Section \ref{sec:discussion}.

No plausible companion candidates are identified in the data. Though a number of sources are visible (see Figure \ref{fig:fomc_with_pm}), they are all either a) likely background sources — based on proper motion versus prior imagery, F356W$-$F444W color, or morphology (e.g., the apparently extended source to the south east) — or b) consistent with residual speckle noise (i.e., inconsistent between the two rolls). Assessment of companion detection limits for these data is provided in Section \ref{sec:companion_limits}. 

\begin{figure*}[h]
\centering
\includegraphics[height=0.37\textwidth]{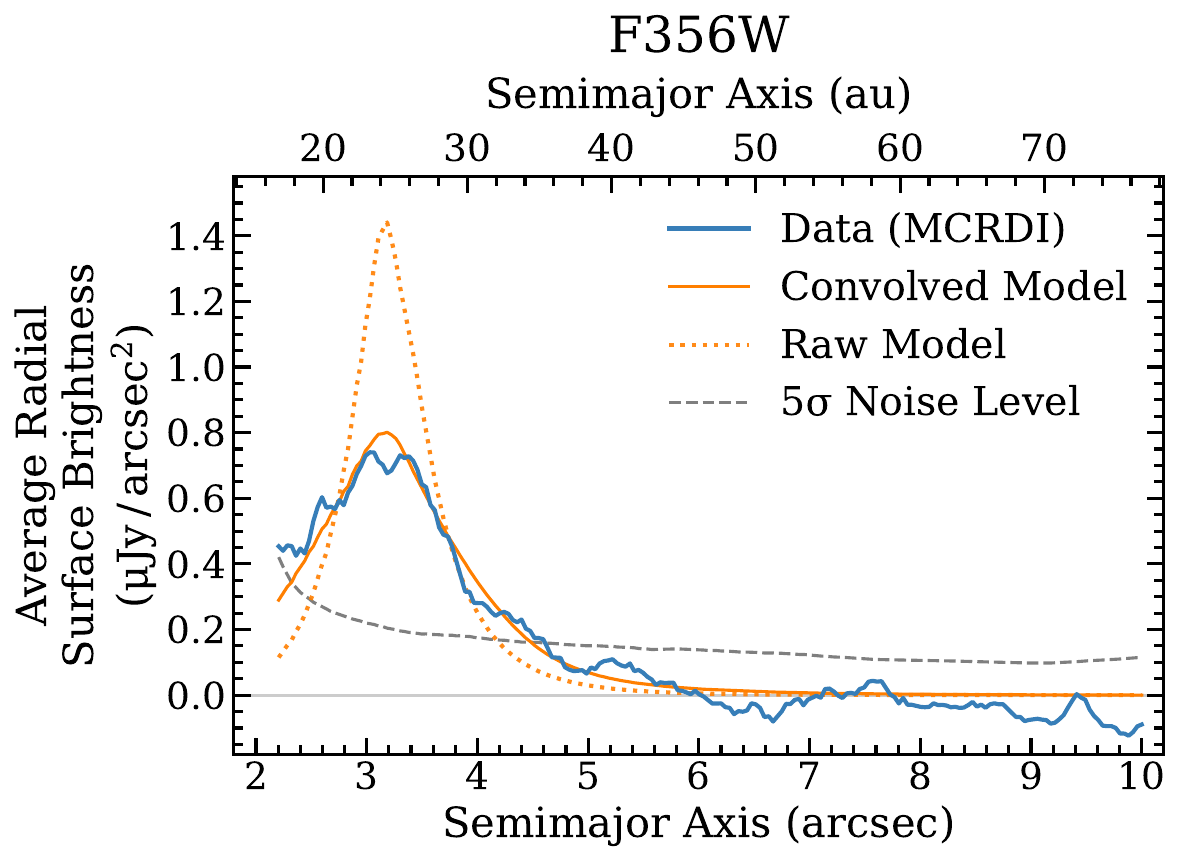}\hspace{1em}\includegraphics[height=0.37\textwidth]{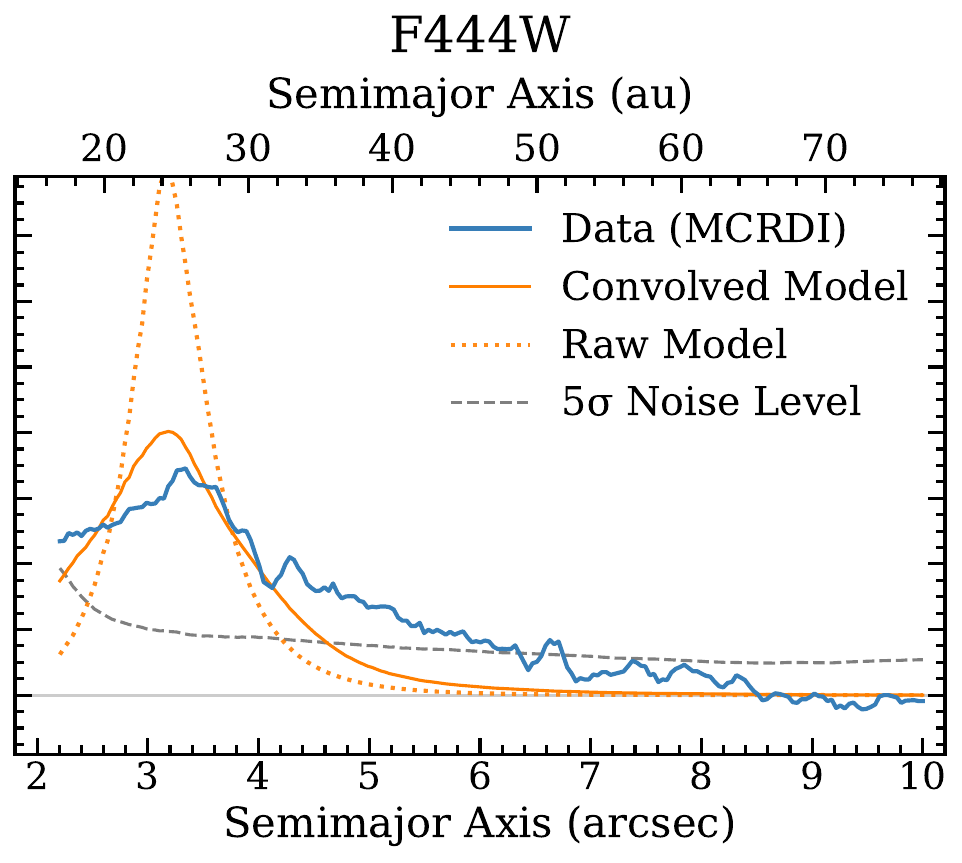}
\caption{Deprojected radial surface brightness profiles for MCRDI reductions of the F356W (left) and F444W (right) data. In each subplot, the profiles for the data, convolved model, and raw model are indicated by the solid blue line, solid orange line, and dotted orange line, respectively. The dashed gray line shows the $5\sigma$ noise level for the data. While the data and convolved model are in good agreement for F356W, for F444W the data shows a deficit at and interior to the disk peak (likely a PSF-subtraction effect) and is significantly more radially extended (discussed in Section \ref{sec:discussion}).\label{fig:radial_sb}}
\end{figure*}

We also provide deprojected radial surface brightness profiles based on the procedure introduced in \citet{Marshall2018} and adopted in \citet{Cronin-Coltsmann2021}, assuming a disk position angle of $-63\degr$ and an inclination of $44\degr$. Measurements for both filters are made in annuli of radial width equal to the instrumental FWHM for F444W (0\farcs17). The deprojected noise level is determined by propagating uncertainties from a radial noise map, generated using the MCRDI reduction with the best-fit disk model subtracted. These profiles are presented in Figure \ref{fig:radial_sb}. Here, the detection in F356W manifests with peak signal to noise ratio of $\sim 17$. These profiles are discussed further in Section \ref{sec:discussion}.

Adopting stellar flux estimates of 387 and 291~mJy for F356W and F444W respectively (see Appendix \ref{app:stellar_flux}), these profiles can be used to estimate the 3--5~\micron{} disk color {(where disk color is the surface brightness color corrected for the color of the incident starlight)}. For this purpose, we adopt the peaks of the raw {(unconvolved)} model curves from Figure \ref{fig:radial_sb} as the disk photometry, as these avoid biasing the measurement by the differences between the filters' PSFs. These values are $1.45\pm0.05$~\mujysb{} at $3.6~\micron{}$ (F356W) and $1.61\pm0.06$~\mujysb{} at $4.4~\micron{}$ (F444W)\footnote{Where uncertainties correspond to the values that increase chi-square by reduced chi-square — excluding less tractable systematic uncertainty; discussed further in Section \ref{sec:discussion}}. {Computing disk color as $\rm S_{3.6}\,/\,S_{4.4} = [F^d_{3.6} \,/\, F^d_{4.4}]\,/\,[F^*_{3.6} \,/\, F^*_{4.4}]$, for disk surface brightnesses $\rm{F}$$^d_\lambda$ and stellar fluxes $\rm{F}$$^*_\lambda$, t}hese measurements correspond to a red disk color of {$\rm S_{3.6}\,/\,S_{4.4} = 0.68\pm0.03$ (or, in magnitudes, $\rm \Delta(m_{3.6} - m_{4.4}) = 0.42\pm0.05$)}. Comparison with both simple Mie theory and the agglomerated debris particle (ADP) models from \citet{Arnold2022}{, each calculated for the F356W and F444W filters,} shows that such a red color is difficult to reproduce for typical size distributions and compositions (e.g., astronomical silicates, water ice, tholins, and olivine). At these wavelengths, we find the reddest colors for either standard astronomical silicates or water ice. For Mie theory models, the observed color is reproduced only for an extremely steep grain size distribution (power law $q\sim-6$), with a minimum grain size of $a_{min} \sim 2.5 ~ \micron$ or $a_{min} \sim 4 ~ \micron$ for compositions of silicates or water ice, respectively; for the ADP models, we identify no solutions reasonably reproducing the measured color. Since dust grains around such low-luminosity stars are not subject to radiation pressure blowout \citep{Arnold2019, Arnold2022}, very small minimum grain sizes are expected — with \citet{Cronin-Coltsmann2022} assuming a value $0.1 ~ \micron$ for Fomalhaut C. As such, a value of $a_{min} \sim 2.5 ~ \micron$ appears implausible, suggesting instead some other cause for the measured color. This result is discussed further in Section \ref{sec:discussion}.

\subsection{Companion Detection Limits}\label{sec:companion_limits}
Before generating contrast curves, we first subtract the best-fitting MCRDI disk model from both the MCRDI and SRDI images to mitigate the impact of the disk on the radial noise estimation. Contrast curves are calculated {as a function of angular separation} using the \texttt{meas\_contrast} routine from the \texttt{PyKLIP} package \citep{Wang2015} for the disk subtracted images and using stellar flux estimates based on TA images (see Appendix \ref{app:stellar_flux}). We then correct these raw contrast curves by dividing them by the coronagraph transmission profile. 

For the MCRDI reductions, we make no additional corrections for PSF subtraction algorithm throughput. While the brightness of any companions would be decreased slightly by the MCRDI procedure \citep[by $\lesssim$ 10\%; see Figure 5 of][]{Lawson2023}, any companion candidates identified with marginal significance could be directly incorporated into the circumstellar model for MCRDI to avoid this \citep[e.g.,][]{Rebollido2024}. As such, the companion itself can be assumed to have no effect on the $5\sigma$ detection limits for MCRDI.

In the case of SRDI, the only comparison made between the PSF model and the science data is when scaling the model brightness to match the data within a narrow annulus spanning 3 to 15 pixels from the star. Any hypothetical companions with flux falling within this annulus would lead to some oversubtraction and would thus have less than 100\% algorithmic throughput. To quantify this, we carry out forward modeling of this effect on synthetic data sequences containing only a companion \ac{psf} (generated using \texttt{WebbPSF}). For both filters, we find that throughput is typically high, $\sim 95\%$, but can dip as low as $\sim 90\%$ for companions that are coincident with brighter stellar speckles and inside the $3<r<15$ pixel annulus. We adopt the median throughput over the sampled position angles at each separation to correct the SRDI contrast curves.

\begin{figure*}
\centering
\includegraphics[width=0.49\textwidth]{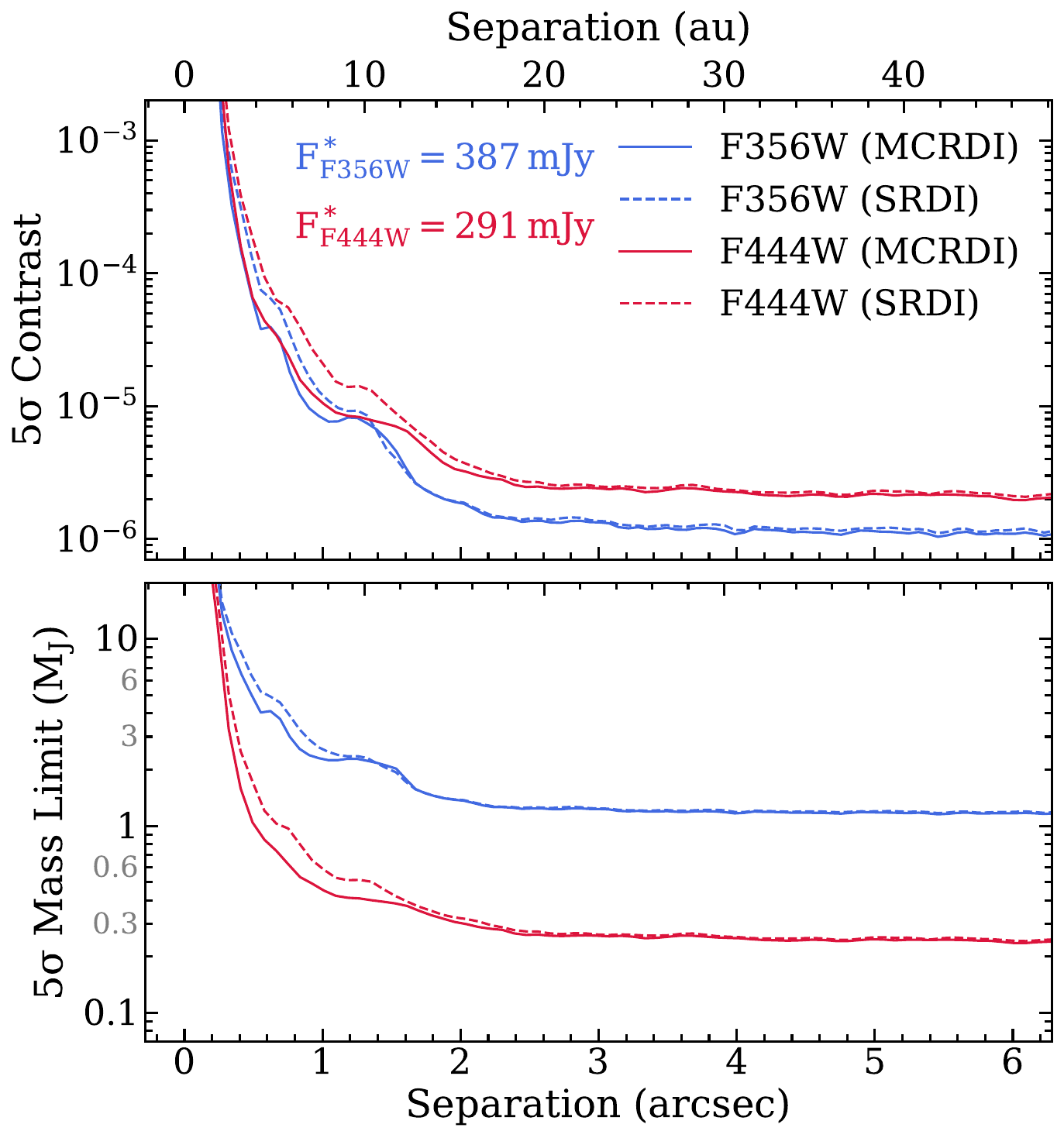}\hspace{1em}\includegraphics[width=0.48\textwidth]{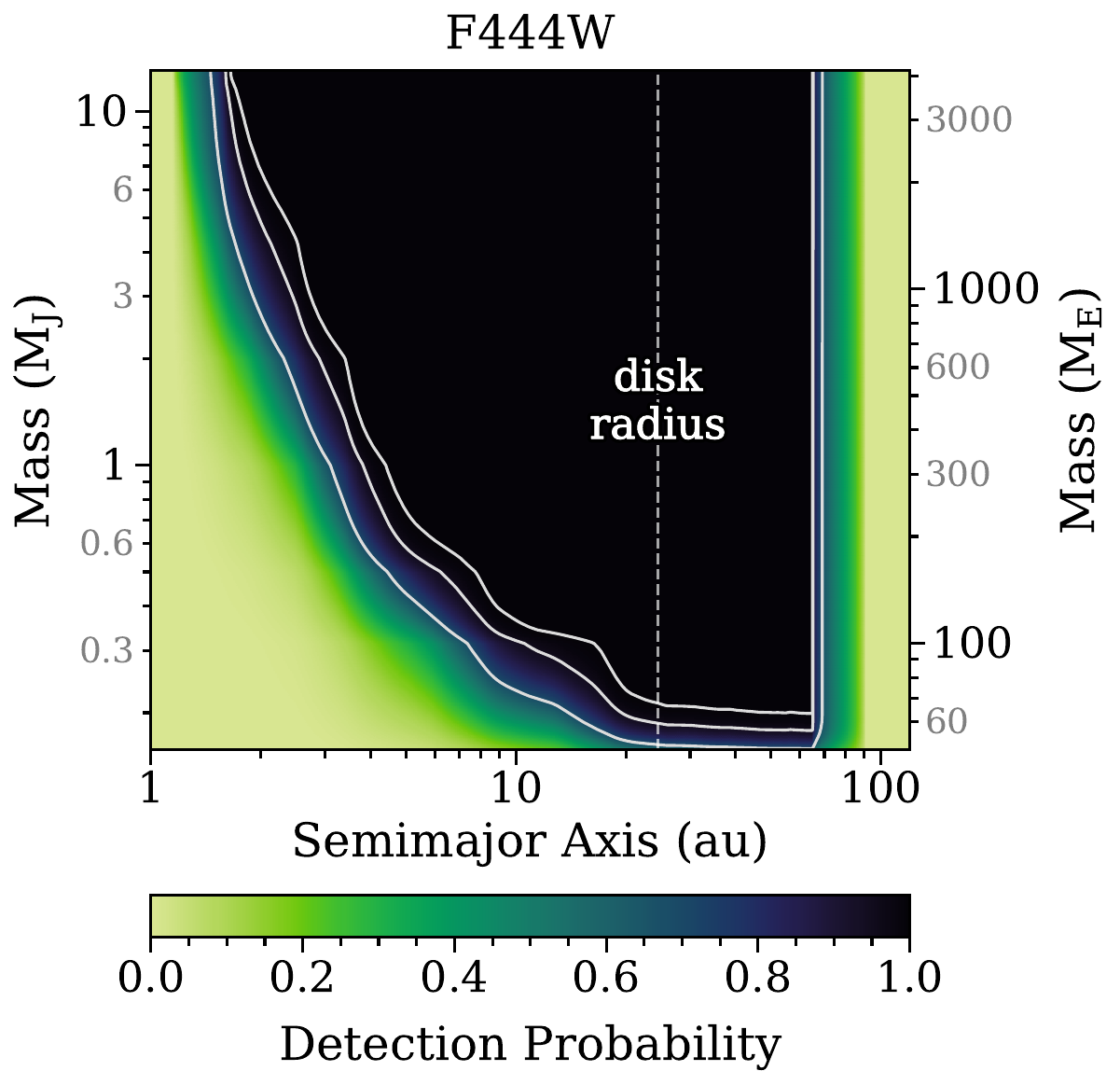}
\caption{Left: Radial $5\sigma$ companion detection limits for model constrained RDI (MCRDI) and synthetic RDI (SRDI) reductions as a function of {angular} separation in arcsec (lower x-axis) and {projected separation in} au (upper x-axis). Limits are provided in terms of contrast with the host star (upper left) and companion mass (lower left). Right: Companion detection probabilities for the F444W (4.4~\micron{}) MCRDI reduction as a function of semimajor axis {(for circular orbits coplanar with the disk)} and companion mass  (Jupiter-masses on the left axis, and Earth-masses on the right axis). White contours mark the {68\%, 95\%, and 99.7\%} probability {thresholds}. For context, the best-fit disk radius of 24.5 au is indicated by a vertical dashed line. These data effectively rule out companions more massive than $\rm \sim 0.3 ~ M_J$ in the vicinity of the disk.\label{fig:contrast}}
\end{figure*}

To map the resulting sensitivities to companion masses, we assume a system age of 440 Myr and adopt synthetic photometry from \citet[][\texttt{BEX-HELIOS} grid]{Linder2019} for companions of $\rm 2~M_{J}$ or less and from \citet[][\texttt{ATMO-CEQ} grid]{Phillips2020} for higher masses. The final detection limits are presented for both filters in the left panel of Figure \ref{fig:contrast}. 

Using these detection limits, we also compute a companion detection probability map for the F444W MCRDI reduction (the most sensitive reduction to low-mass companions){, following the procedure described in \citet{Lawson2023}}. For simplicity, we assume companions on circular orbits oriented in the plane of the disk, with position angles of $-63\degr$ and inclinations of $44\degr$. The resulting probability map is presented in the right panel of Figure \ref{fig:contrast}. These results indicate that the data should have revealed any 0.3$~\rm M_J$ companions beyond $\sim 10$ au, and any 1$~\rm M_J$ companions beyond $\sim 5$ au. These limits are discussed further in Section \ref{sec:discussion}.

\section{Discussion}\label{sec:discussion}
Prior to these results, only four M-dwarf debris disks had been imaged in scattered light: AU Mic, TWA 7, TWA 25, and GSC 07396-00759 \citep{Kalas2004, Choquet2016, Sissa2018}. Fomalhaut C is exceptional among these in having the lowest stellar luminosity by a factor of $\sim 24$ \citep[$\rm L \sim 0.005~L_\odot$;][]{Stassun2019} and the oldest age by a factor of $\sim 18$ \citep[440 Myr;][]{Mamajek2013, Kalas2004, Sissa2018}. As such, Fomalhaut C represents an entirely unprecedented class of scattered-light debris disk host.

The value of these data toward making a more detailed assessment of the disk's scattered-light morphology and composition is limited by the modest significance of the detection and by the presence of numerous background objects and significant residual speckle noise. We can, however, comment tentatively on a number of key parameters in the context of the prior \citet{Cronin-Coltsmann2021} analysis. Modeling of the ALMA detection in \citet{Cronin-Coltsmann2021} identified a particularly narrow best-fitting disk scale width of $\sim$0\farcs11 with a $3\sigma$ upper limit of $0\farcs6$. Meanwhile, the spatial resolution of NIRCam F356W data should be capable of distinguishing scale widths as small as approximately 0\farcs06 (half of the PSF FWHM in the narrowest direction). The best fitting disk model we identify during the MCRDI procedure has a fiducial radius of $\rm r_0 = 24.5~au$ and a radial power law index of $\alpha = 8.6$. This corresponds to a radial scale width of $0\farcs54$ (4.2 au) — just within the reported $3\sigma$ upper limit from \citet{Cronin-Coltsmann2021}. Estimating the uncertainty on the fit parameters via Hessian matrix inversion\footnote{As implemented in the \texttt{LMFIT} package, where parameter uncertainties are taken to be the values increasing chi-square by reduced chi-square; see \href{https://lmfit.github.io/lmfit-py/fitting.html\#uncertainties-in-variable-parameters-and-their-correlations}{``Uncertainties in Variable Parameters, and their Correlations"} for more information.}, yields a $3\sigma$ lower limit on scale width of 0\farcs46. Combined with the smaller disk radius that we identify — $\rm 24.5 \pm 0.1 ~au$ versus $\rm 26.4\pm0.6~au$ from \citet{Cronin-Coltsmann2021} — this could indicate that the scattered light detection corresponds to a population of dust that is not co-located with the material seen by ALMA. We remark, however, that these uncertainties neglect a number of less tractable sources of uncertainty (e.g., the possible presence of unmasked background sources) and thus likely overestimate the significance of these differences. A more confident assessment of this possibility will require higher significance detections.

Unlike the F356W data, the extended emission observed in F444W is not well-explained by a narrow ring-like disk alone. After subtraction of the best-fitting disk model, significant residuals are seen spanning much of the field of view (see Figure \ref{fig:fomc_detections}, lower right panel, and Figure \ref{fig:radial_sb}, right panel). This could be explained in a few ways. First, we may be seeing residual background emission. While our background model assumes a uniform background convolved with the instrumental response, astrophysical variations in the background could produce the observed residuals. In this case, the F444W brightness of the disk derived from forward modeling may also be inflated, which would help to reconcile the exceptionally red color noted in Section \ref{sec:detection}. Alternatively, this could be evidence of an extended outer halo. Some weak evidence of such a halo is seen in analysis of Herschel $160 ~\micron$ PACS data — in the form of an elevated background level \citep{Cronin-Coltsmann2021}. The apparent absence of such a feature in the F356W data could be reconciled if the halo appears more extended at $3.6~\micron{}$ — perhaps due to an increasing presence of smaller grains at wider separations. In this scenario, the halo could appear with a sufficiently shallow slope over the NIRCam field of view so as to be effectively nulled during the background subtraction process. This background oversubtraction would also induce an apparent reddening of the main ring-like component, which would help to explain the seemingly-implausible red disk color that we measure. 

Finally, we note that the departure of the radial profile for the F444W data from that of the model interior to the peak ($\lesssim 3\farcs5$) is consistent with PSF subtraction residuals. Applying PSF subtraction to a simulated disk-free dataset (having the same coronagraph alignments as the real science and reference data) and then repeating these measurements shows a similar negative trend for the F444W at these separations.

The companion detection limits of Section \ref{sec:companion_limits} place strong constraints on the presence of planetary-mass companions around Fomalhaut C. For a mid-M target, the constraints afforded by JWST's sensitivity are unprecedented. While ground-based observations of targets with similar spectral types reach contrasts sufficient to detect masses $\rm \gtrsim 6 ~M_J$ at $2 \arcsec$ \citep[e.g.,][]{Uyama2023}, these JWST NIRCam data reach contrasts sufficient to uncover $\rm \sim 0.3 ~ M_J$ planets at the same separation. These limits can also inform predictions specific to the Fomalhaut C system. \citet{Cronin-Coltsmann2021} note that if the observed radial extent of the disk is the result of material confined between the 3:2 and 2:1 orbital resonances of an unseen planet, then the planet should have a semimajor axis of $\sim 17-20$ au. If such a planet is present in the system, our results effectively constrain its mass to $\rm \lesssim 0.3~M_J$ (approximately the mass of Saturn).

\section{Conclusions}

Based on coronagraphic imagery from JWST~/~NIRCam, we have presented the first scattered-light detections of the Fomalhaut C debris disk — marking the latest spectral type star with a debris disk detected in scattered light to date. We summarize our key findings hereafter.

\begin{enumerate}
\item In both F356W (3.6~$\micron$) and F444W (4.4~$\micron$), the orientation and size of the disk are largely consistent with the prior sub-mm detection with ALMA \citep{Cronin-Coltsmann2021}. In F444W, residual circumstellar flux not detected in F356W could suggest a faint extended halo or localized background variations.
\item By making empirical corrections to synthetic stellar PSFs, we improve contrasts for PSF subtraction with synthetic PSFs by factors of $\sim6$ in the speckle-limited regime and $\sim2$ in the background-limited regime. Compared with other state-of-the-art PSF subtraction methods, this approach achieves comparable contrasts in the background-limited regime, and contrasts within a factor of $\sim 2$ at smaller separations.
\item Modeling of the disk in scattered light favors a slightly smaller disk radius (24.5 au) and broader scale width (0\farcs54) than was identified from the sub-mm ALMA detection. This may indicate that the material probed in scattered light is not co-located with the material seen by ALMA. However, geometric constraints from these data are limited by the significance of the detection and the presence of coincident background sources — such that these modeling results should be considered tentative.
\item No companion candidates are identified; companion detection limits at 4.4~$\micron$ effectively rule out companions more massive than Saturn beyond approximately 10 au and more massive than Jupiter beyond approximately 5 au.
\end{enumerate}

Deeper follow-up observations of Fomalhaut C with JWST/NIRCam would allow better assessment of the disk's scattered-light morphology while more confidently rejecting additional coincident background sources. Combined with improvements to the accuracy of NIRCam target acquisition centroiding since Cycle 1 (Girard et al.~in prep), such observations should also enable a more meaningful probe of the environment interior to the ring as well. This would not only provide clues regarding any yet-unseen companions, but would also better inform the dynamical origins of the Fomalhaut triple system through precise measurement of Fomalhaut C's eccentricity \citep{Shannon2014, Kaib2017, Feng2018, Cronin-Coltsmann2021}.

Overall, these results highlight a key utility for JWST: its ability to study debris disks around the lowest mass stars in scattered light. Future studies leveraging this capability — along with the capability of ALMA to study these elusive disks in thermal emission — stand to make transformational contributions to our understanding of planetary systems around the most common stars. 
\newpage

\acknowledgments
We thank our referee, whose comments helped us to improve both the content and clarity of this manuscript.

We acknowledge the decades of immense effort that enabled the successful launch and commissioning of the JWST; these results were possible only through the concerted determination of thousands of people involved in the JWST mission. In particular, we offer gratitude to a number of individuals who (among others) enabled this study through contributions to either the 2002 NIRCam instrument proposal, the development and commissioning of the NIRCam instrument, or the commissioning of the NIRCam coronagraphy mode: Martha Boyer, Alicia Canipe, Eiichi Egami, Daniel Eisenstein, Bryan Hilbert, Klaus Hodapp, Scott Horner, Doug Kelly, John Krist, Don McCarthy, Karl Misselt, George Rieke, John Stansberry, and Erick Young. We are grateful for support from NASA through the JWST NIRCam project, contract number NAS5-02105 (M. Rieke, University of Arizona, PI). 

The authors thank G. Kennedy for sharing the reduced ALMA data for use in reproducing the \citet{Cronin-Coltsmann2021} contours in our Figure \ref{fig:fomc_with_pm}.

The JWST data presented in this paper were obtained from the Mikulski Archive for Space Telescopes (MAST) at the Space Telescope Science Institute. The specific observations analyzed can be accessed via \dataset[https://doi.org/10.17909/he30-tx49]{https://doi.org/10.17909/he30-tx49}. 
STScI is operated by the Association of Universities for Research in Astronomy, Inc., under NASA contract NAS5–26555. Support to MAST for these data is provided by the NASA Office of Space Science via grant NAG5–7584 and by other grants and contracts.

This publication makes use of data products from the Wide-field Infrared Survey Explorer, which is a joint project of the University of California, Los Angeles, and the Jet Propulsion Laboratory/California Institute of Technology, funded by the National Aeronautics and Space Administration.

This research made use of POPPY, an open-source optical propagation Python package originally developed for the James Webb Space Telescope project (Perrin, 2012).

K. Lawson’s research was supported by an appointment to the NASA Postdoctoral Program at the NASA–Goddard Space Flight Center, administered by Oak Ridge Associated Universities under contract with NASA.

E. Bogat's work was supported by a grant from the Seller's Exoplanet Environments Collaboration (SEEC) at NASA GSFC, administered through NASA's Internal Scientist Funding Model (ISFM). 

\software{
Astropy \citep{astropy2013, astropy2018, astropy2022},
CuPy \citep{cupy2017},
LMFIT \citep{Newville2022},
Matplotlib \citep{matplotlib2007, matplotlib2021},
NumPy \citep{numpy2020},
POPPY \citep{Perrin2012},
pyKLIP \citep{Wang2015},
SciPy \citep{scipy2020},
SpaceKLIP \citep{Kammerer2022},
Vortex Image Processing \citep{Gonzalez2017},
WebbPSF \citep{Perrin2014},
WebbPSF\_ext \citep{Leisenring2021}
}

\appendix

\section{Background Models}\label{app:bg_models}
Treatment of the background in these data requires particular care for a number of reasons. First, the lack of a dedicated reference star is likely to lead to differences in background level between the science and reference images. Typically, a science and dedicated reference target will be near one another on-sky and will be observed in a non-interruptible sequence — leading to comparable backgrounds levels that are ultimately eliminated during PSF subtraction (so long as the stars are also of comparable brightness). Since the targets used as references here were much further on-sky than is typical and sometimes had sizeable offsets in time from the science observations, significantly distinct background levels could manifest. For the observations of Fomalhaut C, the \texttt{jwst\_backgrounds} tool \citep{Rigby2023_jwstbackgrounds} predicts backgrounds of 4.4~\mujysb{} for F356W and 12.4~\mujysb{} for F444W. For the utilized reference target observations, \texttt{jwst\_backgrounds} predicts backgrounds of 2.3–3.4~\mujysb{} for F356W and 5.3–9.5~\mujysb{} for F444W. Second, the background will be affected by the coronagraph's transmission as well as by the neutral density squares in the subarray corners. As such, treating the background as a uniform value (e.g., subtracting a median value from the images) will effectively over-subtract the vicinity of the coronagraph and neutral density squares by differing amounts between images with differing background levels (while under-subtracting elsewhere). This will, in turn, degrade eventual PSF subtraction and likely introduce global photometric inaccuracy. Finally, while these effects may be negligible in scenarios where differences in background levels are substantially smaller than expected sensitivity, the background levels among these targets vary easily on the order of the expected sensitivity and thus cannot be safely ignored. 

Fitting of the background model is carried out prior to alignment of the images, such that the brightness of the convolved background image described in Section \ref{sec:reduction} can simply be scaled in brightness to reproduce the observed background.  In addition to this scaling factor, fitting of the background in these data considers two other components. First, the use of pseudo reference pixels during ramp fitting effectively subtracts a uniform background level from the data. Since the subtracted background level is not known, we simply optimize for a best-fit uniform offset alongside the scalar for the convolved background image (effectively correcting the incidental background subtraction that occurs during ramp fitting). To avoid any contribution from the wings of the stellar PSF, we also subtract a synthetic stellar PSF from the data during the fitting. Since the data have not yet been aligned, cross-correlation is used to align the synthetic PSF with each exposure being considered (as is typically done during image alignment). The overall brightness of the PSF model is then tuned with the other two parameters during the background optimization procedure.

The three parameter model is optimized for each individual exposure (whose integrations are median-combined for this purpose), excluding the region within 60 pixels of the coronagraph center to avoid the portions of the stellar PSF that are poorly described by the cross-correlated PSF model (e.g., due to coronagraph misalignement). We additionally exclude a) a 5 pixel border around the image edge where rows of apparently spurious (significantly positive or negative) values sometimes manifest after ramp fitting, and b) the vicinity of two bright diffraction features that manifest at the subarray edges in the y-axis direction from the target star, but which do not appear in the \texttt{WebbPSF} models. To mitigate the effects of any faint background objects, uncorrected cosmic rays, or detector artifacts, the goodness-of-fit calculation excludes the 10\% most outlying pixels above and below the median of the model-subtracted residuals for each set of trial parameters.

\begin{figure*}
\centering
\includegraphics[width=0.85\textwidth]{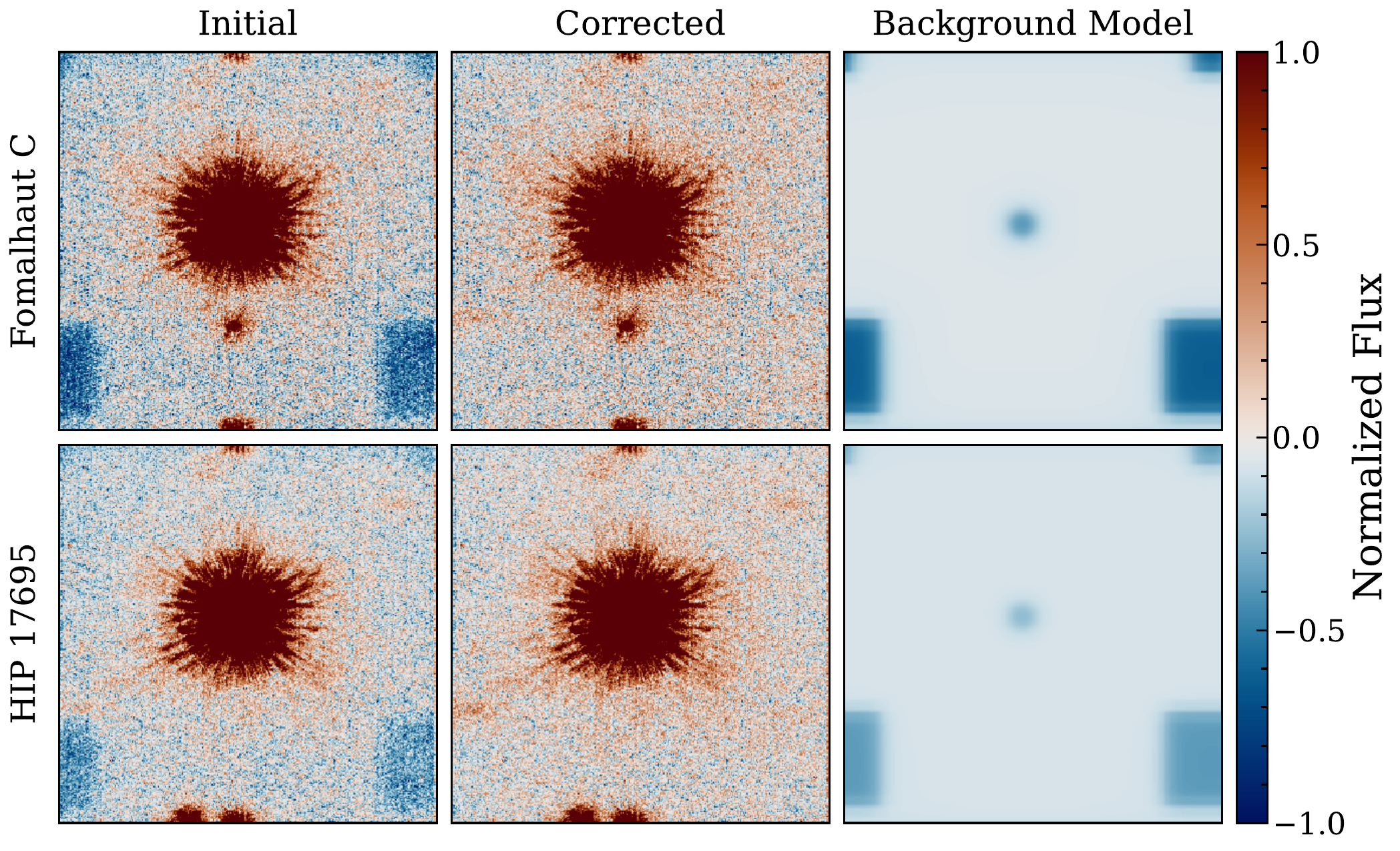}\hspace{1em}
\caption{A comparison of the initial and corrected images from the background model subtraction procedure (Appendix \ref{app:bg_models}) for one science exposure (top row) and one reference exposure (bottom row) in F356W. The images of each row are normalized to the initial stellar PSF brightness to highlight the difference between the backgrounds (visible as initially more negative regions within the neutral density squares at the corners). The right-most column shows the background model that is subtracted from the initial image. \label{fig:background_model}}. 
\end{figure*}

Once the optimization is completed for each exposure, we add the best-fit value for the uniform offset back to each individual integration, and then subtract the best-fit convolved background image to reach the final background-subtracted images. Figure \ref{fig:background_model} shows the results of this procedure for both a science and a reference exposure.

\section{MCRDI Subtraction Details}\label{app:mcrdi_details}

The initial unconstrained RDI reduction for each filter considers two annular optimization regions spanning stellocentric separations of $4 \leq r < 25$ pixels and $25 \leq r < 100$ pixels, with corresponding subtraction regions of $r < 25$ pixels and $r \geq 25$ pixels, respectively. The optimization regions additionally exclude the vicinity of four off-axis (likely background) sources in the science target's field of view. The radial boundaries for these regions were selected to approximately isolate the inner speckle-dominated regime. At larger separations, a total of 13 reference integrations featured significant uncorrected cosmic ray halos. By splitting the subtraction into two regions, we were able to exclude the affected frames for only the outer region, while allowing the inner region to benefit from a larger reference library. For each subtraction region, the PSF model is constructed as the linear combination of the reference images that minimizes squared residuals with the science image within the corresponding optimization region. 

Following the initial RDI procedure, disk forward modeling is carried out. Raw disk models are generated using a simplified version of the GRaTer code \citep{Augereau1999} implemented in \texttt{Vortex Image Processing} \citep[\texttt{VIP};][]{Gonzalez2017}. For the scattering phase function (SPF) of the disk, we adopt a simple Henyey-Greenstein (H-G) SPF \citep{Henyey1941}. Though such an SPF is not expected to be physically informative \citep[e.g.,][]{Hughes2018}, the primary intent of the disk model here is to superficially reproduce the observed disk in order to suppress RDI oversubtraction. For this purpose, the adopted SPF is suitable.

The disk's inclination and position angle are fixed to the values from the overall best model reported in \citet{Cronin-Coltsmann2021}: $44\degr$ and $-63\degr$ respectively. The varied parameters for the F356W model include fiducial radius, scale height, density power-law index ($\alpha = \alpha_{in} = -\alpha_{out}$), H-G asymmetry parameter, eccentricity, and argument of periapsis. Additionally, an optimal overall brightness scaling factor is determined analytically for each model (i.e., this parameter is not varied by the optimizer). For the F444W model, we fix the parameters governing the disk's geometry to the best-fit values from the F356W procedure — tuning only the asymmetry parameter and the overall brightness scalar. 

Convolution of disk models follows the procedure of \citet{Lawson2023}, using a grid of synthetic PSF models and coronagraph transmission maps generated with consideration for the mask position in each roll after image alignment. The PSF grid samples the origin as well as 12 logarithmically-spaced radial positions at each of four linearly-spaced azimuthal positions (for a total of 49 spatial samples). Each model is then forward-modeled as typical for RDI, evaluating goodness of fit with a simple $\chi^2$ metric within the outer optimization region described above (to mitigate the impact of residual speckle noise in the inner region).

After the best-fit model is identified, that model is used to carry out an MCRDI reduction of the data. In this process, the stellar PSF model is constructed by comparing the reference images with science images from which the optimal disk model has been subtracted. The resulting stellar PSF model is then subtracted from the original science frames. Besides the disk model ``constraint", the PSF subtraction for the MCRDI reduction is identical to the initial unconstrained RDI procedure.

\section{Empirical Corrections for Synthetic RDI}\label{app:synth_rdi}

Empirical corrections to the nominal synthetic reference PSF model for the science data are determined as follows. For each reference exposure, we generate a nominal PSF model image using \texttt{WebbPSF} as we did for Fomalhaut C. The brightness of each PSF model is scaled to minimize squared residuals with the data within an annular region spanning $3\leq r < 15$ pixels from the star. The model is then subtracted from the data and the residuals are divided by the total model flux to produce a normalized residual map for each reference exposure. 

Inspection of the residual maps shows that the model inaccuracy changes little between exposures. This suggests that the bulk of the PSF model inaccuracy results from characteristics that do not change across the sample of reference exposures — such as underlying simplifications in the \texttt{WebbPSF} optical model. The most significant changes that do occur in the residuals from exposure to exposure appear to be temporally correlated; for example, the residual maps for AP Col and 2MJ0944 (November 2022) are more similar to one-another than to those of G-7-34 and HIP 17695 (October 2022). Temporally correlated errors could occur as a result of differences in OPD file accuracy. As the data are distributed roughly into two groups of observations — October 2022 and November 2022 — such a correlation could manifest if a tilt event occurred between the time of one set of observations and the utilized OPD file, resulting in a less accurate OPD measurement for those exposures. However, comparison of the OPD measurements flanking each observation show no evidence of such an event — with $\rm \Delta WFE = 6.4~$nm RMS between the preceding and succeeding OPD files for both the October and November data — below the Cycle 1 median of 9~nm RMS \citep{Lajoie2023}.

Since the GTO 1184 program was observed roughly in order of spectral type, the underlying driver of this phenomenon could also be related to the spectra of the targets. For example, synthetic PSF inaccuracies correlated with spectral type could manifest if a) the utilized synthetic spectra are inaccurate as a function of spectral type, or b) the inaccuracies in the WebbPSF models have wavelength dependence. Ultimately, these data are insufficient to confidently disentangle these possibilities. Conducting this analysis on a broader and more diverse set of observations is necessary to better understand the relevant factors.

\begin{figure*}
\centering
\includegraphics[width=0.36\textwidth]{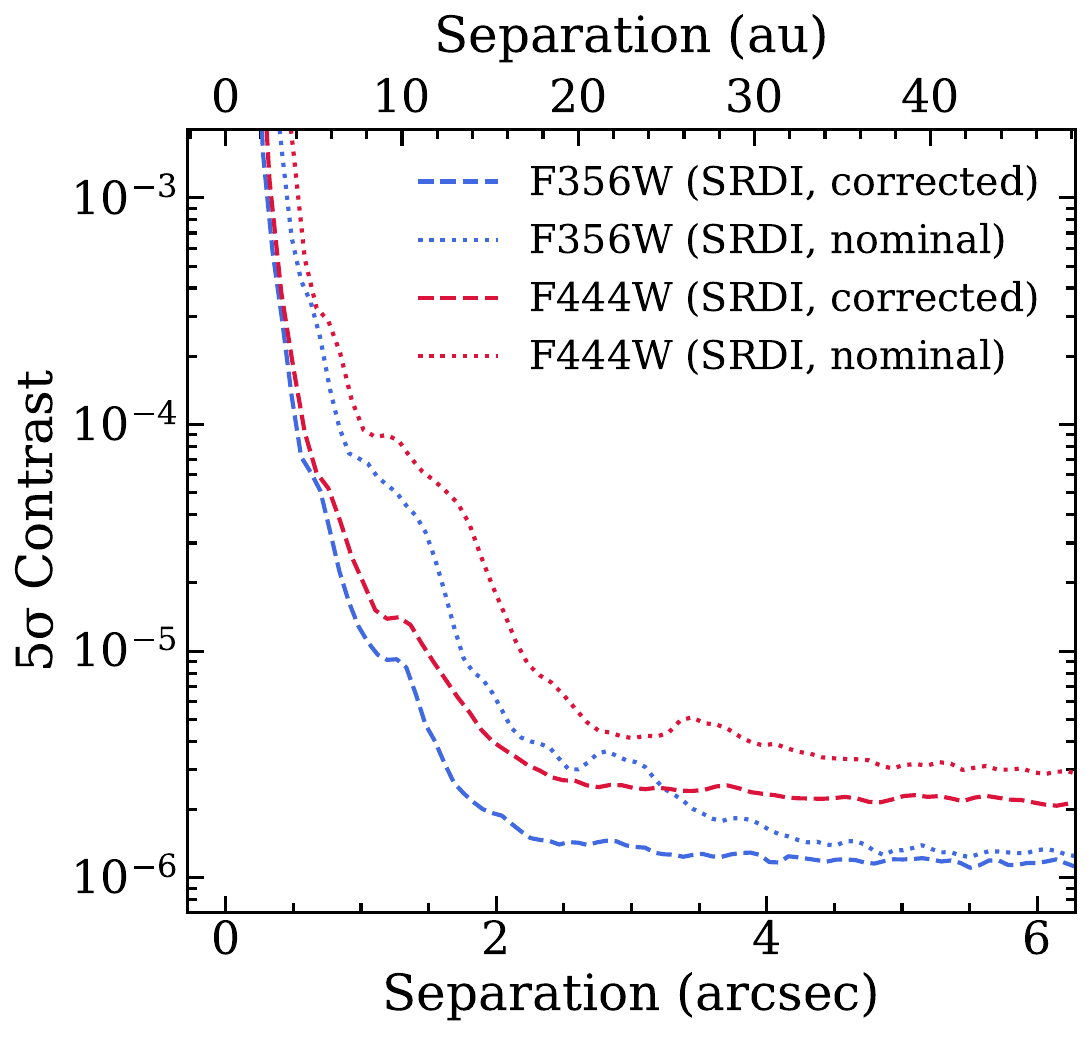}\hspace{1em}
\includegraphics[width=0.6\textwidth]{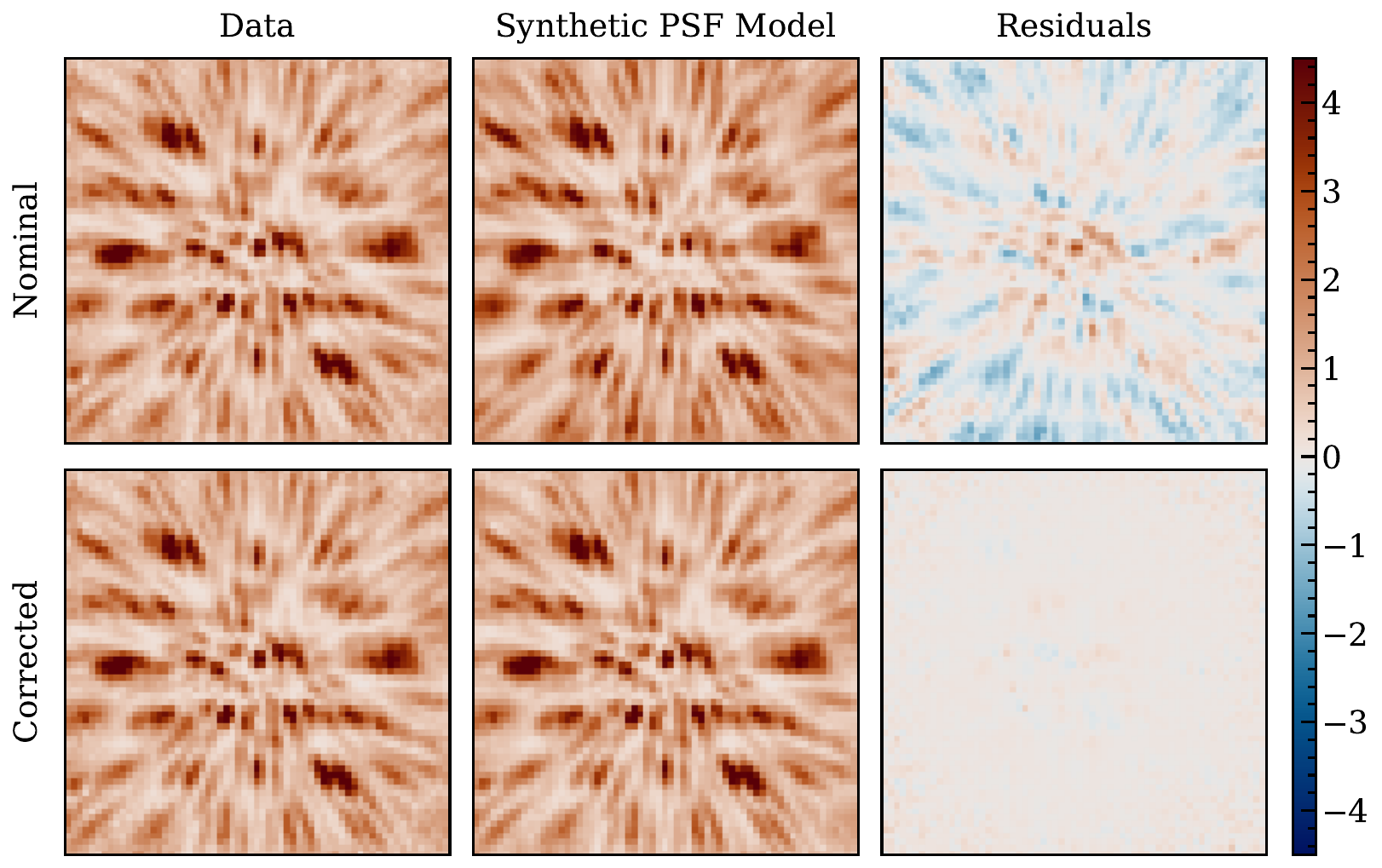}
\caption{Left: Contrast curves for the nominal (dotted lines) and corrected (dashed lines) synthetic RDI reductions of the Fomalhaut C data. The implemented corrections improve contrasts by up to a factor of six for both filters. Right: A comparison of synthetic PSF models for one roll of Fomalhaut C F356W data without (top row) and with (bottom row) the empirical corrections described in Appendix \ref{app:synth_rdi}. All images were divided by the radial average for the data to better compare the speckle patterns at all separations. The field of view for each panel is $2\times2\arcsec$. \label{fig:synth_rdi}}. 
\end{figure*}

To determine the model correction map for Fomalhaut C, we take the exposure-wise median of the normalized residual maps for exposures of both AP Col and 2MJ0944. These targets are selected because they were observed closest in time to the science target, and should thus yield the most accurate correction based on the analysis above. To produce the final corrected PSF model for each science exposure, we multiply the normalized correction map by the total science model flux and add this to the nominal PSF model. In Figure \ref{fig:synth_rdi}, we compare the resulting contrasts between the nominal and corrected synthetic \ac{rdi} reductions and provide example images for one roll of the F356W data. The apparently worse performance for SRDI in F444W than in F356W (see Figure \ref{fig:fomc_detections}) could be evidence that the nominal PSF model deficiencies are indeed spectrally dependent.

The cost of this procedure is the introduction of noise to the otherwise noiseless synthetic PSF model. For this particular application, the noise introduced this way is significantly outweighed by the improved suppression of the diffraction pattern at all separations. In broader applications, it may be beneficial to adopt model corrections only at separations where this balance is favorable.

\section{NIRCam Stellar Flux Estimation}\label{app:stellar_flux}
To enable assessment of the sensitivity of these data in terms of contrast, we approximate the flux of Fomalhaut C in each filter as follows. We begin with the unocculted target acquisition images (one for each roll), which were taken in the F335M filter (3.35~\micron{}, no neutral density square), and convert them from units of MJy/sr to mJy/pixel$^2$. We measure initial fluxes for each image as the sum of pixels within a 15 pixel radius aperture centered on the PSF peak. To make aperture corrections to these fluxes, we use \texttt{WebbPSF} to compute the encircled energy at 15 pixels for a synthetic PSF normalized such that an infinite aperture would measure a total flux of 1 at the exit pupil. We divide each initial flux measurement by this encircled energy fraction, yielding F335M fluxes of 419 and 433 mJy. We adopt the average of these, 426 mJy, as the F335M flux. Finally, we scale an approximate synthetic AMES-Cond spectrum for Fomalhaut C to produce the measured F335M flux and then extract F356W and F444W fluxes from the result. This yields F356W and F444W fluxes of 387 mJy and 291 mJy, respectively.

\bibliography{refs}{}
\bibliographystyle{aasjournal}
\end{document}

%% file: acronyms.tex
\newacro{adi}[ADI]{angular differential imaging}
\newacro{ao}[AO]{adaptive optics}
\newacro{aolp}[AOLP]{angle of linear polarization}
\newacro{bff}[BFF]{Best Factor Finding}
\newacro{charis}[CHARIS]{Coronagraphic High Angular Resolution Imaging Spectrograph}
\newacro{css}[CSS]{circumstellar signal}
\newacro{de}[DE]{differential evolution}
\newacro{disnmf}[DI-sNMF]{data imputation using sequential nonnegative matrix factorization}
\newacro{dpp}[DPP]{Data Processing Pipeline}
\newacro{drp}[DRP]{Data Reduction Pipeline}
\newacro{fov}[FOV]{field of view}
\newacro{fwhm}[FWHM]{full width at half maximum}
\newacro{gpi}[GPI]{Gemini Planet Imager}
\newacro{gto}[GTO]{Guaranteed Time Observations}
\newacro{hiciao}[HiCIAO]{High-Contrast Coronographic Imager for Adaptive Optics}
\newacro{hst}[HST]{Hubble Space Telescope}
\newacro{hwp}[HWP]{half-wave plate}
\newacro{i}[I]{total intensity}
\newacro{ir}[IR]{infrared}
\newacro{ifs}[IFS]{integral field spectrograph}
\newacro{irdis}[IRDIS]{InfraRed Dual-band Imager and Spectrograph}
\newacro{iwa}[IWA]{inner working angle}
\newacro{jwst}[JWST]{the James Webb Space Telescope}
\newacro{klip}[KLIP]{Karhunen-Lo\`{e}ve Image Projection}
\newacro{loci}[LOCI]{Locally Optimized Combination of Images}
\newacro{magaox}[MagAO-X]{MagAO-X}
\newacro{mloci}[MLOCI]{Matched LOCI}
\newacro{nir}[NIR]{near-infrared}
\newacro{nircam}[NIRCam]{Near Infrared Camera}
\newacro{nmf}[NMF]{Non-negative Matrix Factorization}
\newacro{opd}[OPD]{optical path difference}
\newacro{pca}[PCA]{principal component analysis}
\newacro{pdi}[PDI]{polarimetric differential imaging}
\newacro{pi}[PI]{polarized intensity}
\newacro{piaacmc}[PIAACMC]{phase-induced amplitude apodization complex mask coronagraph}
\newacro{ppd}[PPD]{protoplanetary disk}
\newacro{psf}[PSF]{point spread function}
\newacro{rdi}[RDI]{reference star differential imaging}
\newacro{sb}[SB]{surface brightness}
\newacro{scexao}[SCExAO]{Subaru  Coronagraphic  Extreme  Adaptive  Optics}
\newacro{sdi}[SDI]{spectral differential imaging}
\newacro{sed}[SED]{spectral energy distribution}
\newacro{snr}[SNR]{signal-to-noise ratio}
\newacro{snre}[SNRE]{signal-to-noise per resolution element}
\newacro{spf}[SPF]{scattering phase function}
\newacro{sphere}[SPHERE]{Spectro-Polarimetric High-contrast Exoplanet REsearch instrument}
\newacro{vampires}[VAMPIRES]{Visible Aperture Masking Polarimetric Imager for Resolved Exoplanetary Structures}
\newacro{vapp}[vAPP]{vector Apodizing Phase Plate}
\newacro{vlt}[VLT]{Very Large Telescope}